\documentclass[fleqn,usenatbib]{mnras}

\usepackage{newtxtext,newtxmath}
\usepackage[T1]{fontenc}

\DeclareRobustCommand{\VAN}[3]{#2}
\let\VANthebibliography\thebibliography
\def\thebibliography{\DeclareRobustCommand{\VAN}[3]{##3}\VANthebibliography}

\usepackage{graphicx}	
    \graphicspath{{./}{figures/}}
\usepackage{amsmath}	
\usepackage{siunitx}

\newcommand{\dr}{$\Delta\!R$}
\newcommand{\Dlb}{$\Delta\!L\!B$}
\newcommand{\hp}{half-power method}
\newcommand{\ip}{inflection-point method}
\newcommand{\lb}{$L\!B$}
\newcommand{\lbc}{$L\!B$\textsubscript{conv}}
\newcommand{\lbp}{$L\!B$\textsubscript{synth}}
\newcommand{\BW}{{\small HPBW}}
\newcommand{\hpN}{{\small HPBW}${}_0$}
\newcommand{\Rb}{$\bar{R}$}
\newcommand{\Rc}{\textit{R}\textsubscript{conv}}
\newcommand{\Rp}{\textit{R}\textsubscript{synth}}
\newcommand{\Rt}{$R_\theta$}
\newcommand{\Rssc}{\textit{R}\textsubscript{ssc}}
\newcommand{\RsscC}{$R_\mathrm{ssc}^\mathrm{conv}$}
\newcommand{\tb}{$T\!_B$}
\newcommand{\tqs}{$T\!_{qS}$}
\newcommand{\tbqs}{\tb$/$\tqs$-1$}
\newcommand{\alma}{{\small ALMA}}
\newcommand{\sst}{{\small SST}}
\newcommand{\hs}{$\;\;\;\;\;\;$}
\let\mc\multicolumn

\title{Subterahertz Radius and Limb Brightening of the Sun Derived from SST and ALMA}
\author[F. Menezes et al.]{
Fabian Menezes,$^{1}$\thanks{E-mail: menezes.astroph@gmail.com}
Caius L. Selhorst,$^{2}$
Carlos Guillermo Giménez de Castro$^{1,3}$
and Adriana Valio$^{1}$ 
\\
$^{1}$Centro de R{\'a}dio Astronomia e Astrof{\'i}sica Mackenzie (CRAAM), Universidade Presbiteriana Mackenzie, S{\~a}o Paulo, Brazil\\
$^{2}$N{\'u}cleo de Astrof{\'\i}sica, Universidade Cruzeiro do Sul / Universidade Cidade de S{\~a}o Paulo, S{\~a}o Paulo, SP,  Brazil\\
$^{3}$Instituto de Astronomía y Física del Espacio, UBA/CONICET, Buenos Aires, Argentina.
}

\date{Accepted 2021 November 28. Received 2021 November 23; in original form 2021 March 26}

\pubyear{2021}

\begin{document}
\label{firstpage}
\pagerange{\pageref{firstpage}--\pageref{lastpage}}
\maketitle

\begin{abstract}
Measurements of the radius and limb brightening 
of the Sun provide important information about the solar atmosphere structure and temperature. 
The solar radius 
increases as the observation at radio frequency decreases, indicating that each emission originates higher in the  atmosphere. Thus, different layers of the solar atmosphere can be probed by observing at multiple wavelengths. In this work, we determined the average radius and limb brightening at 100, 212, 230, and 405 GHz, 
using data from the Solar Submillimeter Telescope 
and ALMA's single-dish observations. For the first time, limb brightening values for frequencies of 212 and 405 GHz were estimated. At sub-THz frequencies, the observed limb brightening 
may affect the solar radius measurements. We use two different and well known approaches to determine the radius: the \hp\ and the \ip. 
We investigate how the antenna beam size 
and the limb brightening level, \lb, can affect the radius measurements using both methods. Our results showed that the inflection-point method is the least affected by these parameters, and should thus be used for solar radius estimates at radio wavelengths.  The 
measured average radii are 
$\ang{;;968}\pm\ang{;;3}$ (100 GHz), $\ang{;;963}\pm\ang{;;3}$ (212 GHz), $\ang{;;963}\pm\ang{;;2}$ (230 GHz), and $\ang{;;963}\pm\ang{;;5}$ (405 GHz).
Finally, we used forward modeling to estimate the ranges of \lb\ of the solar disk resulting in
5\%-19\% (100 GHz), 2\%-12\% (212 GHz), 6\%-18\% (230 GHz), and 3\%-17\% (405 GHz). Both radius and  limb brightening estimates
agree with previous measurements reported in the literature. 
\end{abstract}

\begin{keywords}
Sun: atmosphere -- Sun: radio radiation -- Sun: chromosphere -- methods: data analysis
\end{keywords}



\section{Introduction} \label{sec:intro}

Investigated and measured for over 2 millenia, the optical radius of the Sun is a fundamental parameter. Records of measurements by Greek astronomers date back to the 3\textsuperscript{rd} century B.C., when they determined the apparent mean solar radius between \ang{;;810} and \ang{;;988} \citep{rozelot18}. Despite their ingenuity, the measurements were highly uncertain, and it took centuries for this matter to be revisited, regaining attention only in the 17\textsuperscript{th} century and being often studied from then on. \citet{gilliland81} and \citet{vaquero16}, for instance, compiled data from the 18\textsuperscript{th} until the 21\textsuperscript{st} century obtained from different observatories and authors. With measurements of the solar radius, other features and aspects were discovered, such as the variation of the solar apparent size throughout the year due to Earth's orbital eccentricity. Also, measurements of small variations in the solar optical radius are a critical probe of the Sun's interior stratification \citep{emilio00}. 

Due to technological limitations, until some decades ago only optical measurements were available. Since then, measurements at radio frequencies began to take place \citep{coates58}. By observing the Sun at different radio wavelengths, the radius measurements present an exponential trend, as showed by \citet{menezes21}, decreasing with frequency. Therefore at a specific wavelength, the difference of the measured radius from the optical radius can be understood as the height above the photosphere where most of the emission is being created in the limb regions. Moreover, the study of the solar radius at different wavelengths, derived from eclipse and direct observations, enables probing the solar atmosphere throughout the solar activity cycle \citep{swanson73, costa99, menezes17, selhor19}.

The canonical value at optical wavelengths of the mean apparent solar radius is \ang{;;959.63} \citep{auwers1891}. There are different techniques to measure the solar radius, such as helioseismic methods from the $f$-modes \citep{antia98, kosovichev18a}, determination from total solar eclipse observations \citep{kubo93, kilcik09}, and from direct observations \citep{emilio00}. \citet{brown98} used the Solar Diameter Monitor instrument \citep{brown82} and defined the solar limb using the finite Fourier transform definition (FFTD) to measure the optical radius.

At radio wavelengths, the radius can be measured by the \ip\ \citep{alissan17, menezes21}, and by the \hp\ \citep{costa99, selhor11, menezes17}. The measurement of the solar radius by any of theses two methods is considerably affected by the presence of limb brightening in the solar disk. Mainly, because the center-to-limb average distance of the solar disk is used to define the radius, and the limb coordinates determination is affected by the limb brightening level, \lb, of the brightness temperature, \tb, profile. In this work we define $L\!B = T_B / T_{qS} - 1$, \textit{i.e.}, the percentage above the quiet Sun temperature, \tqs.

The solar limb brightening (or darkening) is a property that provides important information about the temperature gradient of the solar atmosphere. Solar atmospheric models predict a positive temperature gradient above the photospheric minimum temperature \citep[e.g.,][]{vernazza81, fontenla93, selhor05}. This translates to an increase in the \tb\ from submillimetric to centimetric wavelengths, because the emission at these wavelengths are formed above the minimum temperature region, where the temperature gradient is positive.

Based on maps obtained since 1992 by the Nobeyama Radioheliograph at 17 GHz \citep[NoRH,][]{nakajima94}, \cite{shibasaki98} reported regions near the solar poles with \tb\ values up to 40\% above the quiet-Sun value, in contrast with equatorial limb regions which presented only a 10\% increase. The limb brightening of the solar disk has been registered in observations at several bands of the electromagnetic spectrum: at 1.3 mm \citep{withbroe70}, UV \citep{horne81}, between 350 $\mu$m and 1 m \citep{lindsey76}, at 17 GHz \citep{selhor03, shibasaki98}, 860 GHz \citep{lindsey81}, and at 100 and 230 GHz \citep{selhor19}. According to \citet{selhor03}, at 17 GHz the average intensity of the poles was found to be approximately 13\% and 14\% above the quiet-Sun level, \tqs, in the North and South poles, respectively. At 100 GHz, \cite{selhor19} found that the average brightness above \tqs\ was 6.1\% $\pm$ 2.8\% at the North pole and 5.1\% $\pm$ 2.7\% at the South pole; whereas at 230 GHz, the average values were 9\% $\pm$ 5\% and 9\% $\pm$ 4\%, respectively for the South and North poles.

At sub-THz frequencies -- \textit{i.e.}, millimeter and submillimeter wavelengths -- there is a lack of measurements of the radius, as well as other parameters of the solar atmosphere. Since observation at distinct wavelengths probe different layers of the solar atmosphere, the study of the solar radius and limb brightening provides important information about the solar atmosphere structure and temperature.

This work is a follow-up of \cite{menezes21}, in which we measured the average equatorial and polar radius at 212  and 405 GHz (using all receivers of the Submillimeter Solar Telescope) from 2007 to 2019, and at 100 and 230 GHz from 2015 to 2018. Also it was focused on the relation between solar radius time series and solar activity. Here, we focus on the mean radius and limb of the full sub-THz solar disk. We measure the mean solar radii at 100, 212 (receiver \#3), 230, and 405 GHz (receiver \#5) during a 4-year, period from 2015 to 2018. Two methods to calculate the radius at radio frequencies are considered and compared: the inflection-point and the half-power (defined in Section~\ref{sec:rad_determ}). Forward modeling is used to investigate the effect that parameters such as antenna beam size (\BW) and \lb\ have on the solar radius measurements for both methods. Finally, we estimate the limb brightening above quiet-Sun values for these sub-THz frequencies.

\section{Observations} \label{sec:method}

\subsection{Radio Telescopes} \label{sec:instr}

For the determination of the solar radii and the estimate of the limb brightening we used data provided by solar observations of the Atacama Large Millimeter/submillimeter Array \citep[ALMA;][]{wootten09} at 100 and 230 GHz, and by the Solar Submillimeter-wave Telescope \citep[SST;][]{kaufmann08} at 212 and 405 GHz. ALMA is an international radio interferometer located in the Atacama Desert of northern Chile at 5000 m elevation.
 
SST is a radio telescope located at CASLEO Observatory in Argentina, at 2552 m elevation. The telescope is composed of six receivers, being two radiometers at 405 GHz and four at 212 GHz, with an absolute pointing accuracy of about \ang{;;10}. Since structural antenna deformations change the expected size and form of the beams \citep{valle21}, in this study we only use receivers \#3 (212 GHz) and \#5 (405 GHz), which beams have less deformation compared to the other ones. 

\subsection{Solar maps}

SST solar maps are obtained from parallel or radial raster scans of the Sun covering an area of about a $\ang{1;;}\times\ang{1;;}$ in the sky with a tracking speed in the range of \ang{0.1;;} and \ang{0.2;;} per second. A few solar maps are reconstructed from radial and RA-dec\footnote{RA-dec: right ascension and declination} raster scans, whereas the vast majority of the solar maps are reconstructed from azimuth-elevation scans with a \ang{;2;} separation between scans obtained \citep{gimenez20, valle21, menezes21}. For a typical integration time of 0.04s, the rectangular pixels are $\ang{;0.48;}\times\ang{;2;}$ in size. The SST maps were then interpolated to square maps of $600\times600$ pixels.

\begin{figure}
\centering
\includegraphics[width=1.0\columnwidth]{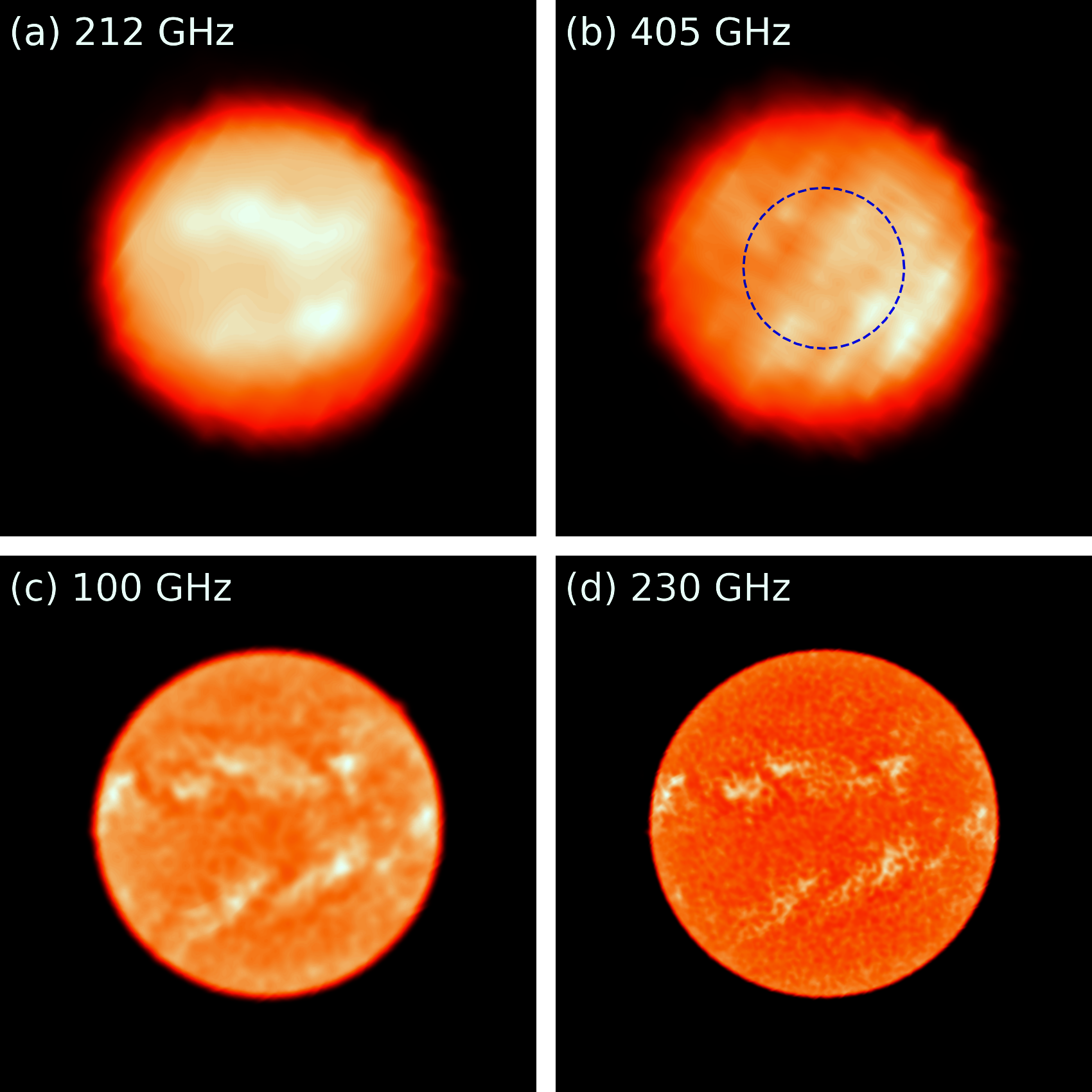}
\caption{Solar maps obtained on 2015-12-17 from SST (upper panels) and ALMA (lower panels) at (a) 212 GHz, (b) 405 GHz, (c) 100 GHz, and (d) 230 GHz. The blue circle (b) with a radius of \ang {;7.5;} constrains the area used to define \tqs.}
\label{fig:maps} \end{figure}

During the years of 2015 to 2018, a total of 2148 maps at 212 GHz and 1256 maps at 405 GHz were selected for this study (the selection criteria are described in Section \ref{sec:rad_determ}). Examples of the SST azimuth-elevation interpolated maps at 212 GHz (top left panel) and 405 GHz (top right panel) are shown in Fig.~\ref{fig:maps}, where the color scale represents \tb.

From ALMA, we use fast-scan single-dish maps obtained in Band 3 (84–116 GHz, hereafter, 100 GHz) and in Band 6 (211–275 GHz, hereafter, 230 GHz), described by \citet{white17}. We analyzed 125 maps at 100 GHz and 71 at 230 GHz (196 in total). These maps were made during four solar observation campaigns from 2015 to 2018 by ALMA's 12-meter single-dish. The images were extracted from FITS files provided by ALMA's science verification page and ALMA science archive page\footnote{URLs are provided in the Data Availability section, and ALMA project codes in the Acknowledgments section}. Examples of ALMA maps obtained at 100 GHz (bottom left panel) and 230 GHz (bottom right panel) are shown in Fig.~\ref{fig:maps}, where the color scale represents the brightness temperature, \tb.

\subsection{Antenna beam} \label{sec:beam}

\begin{figure*}
\includegraphics[width=1.4\columnwidth]{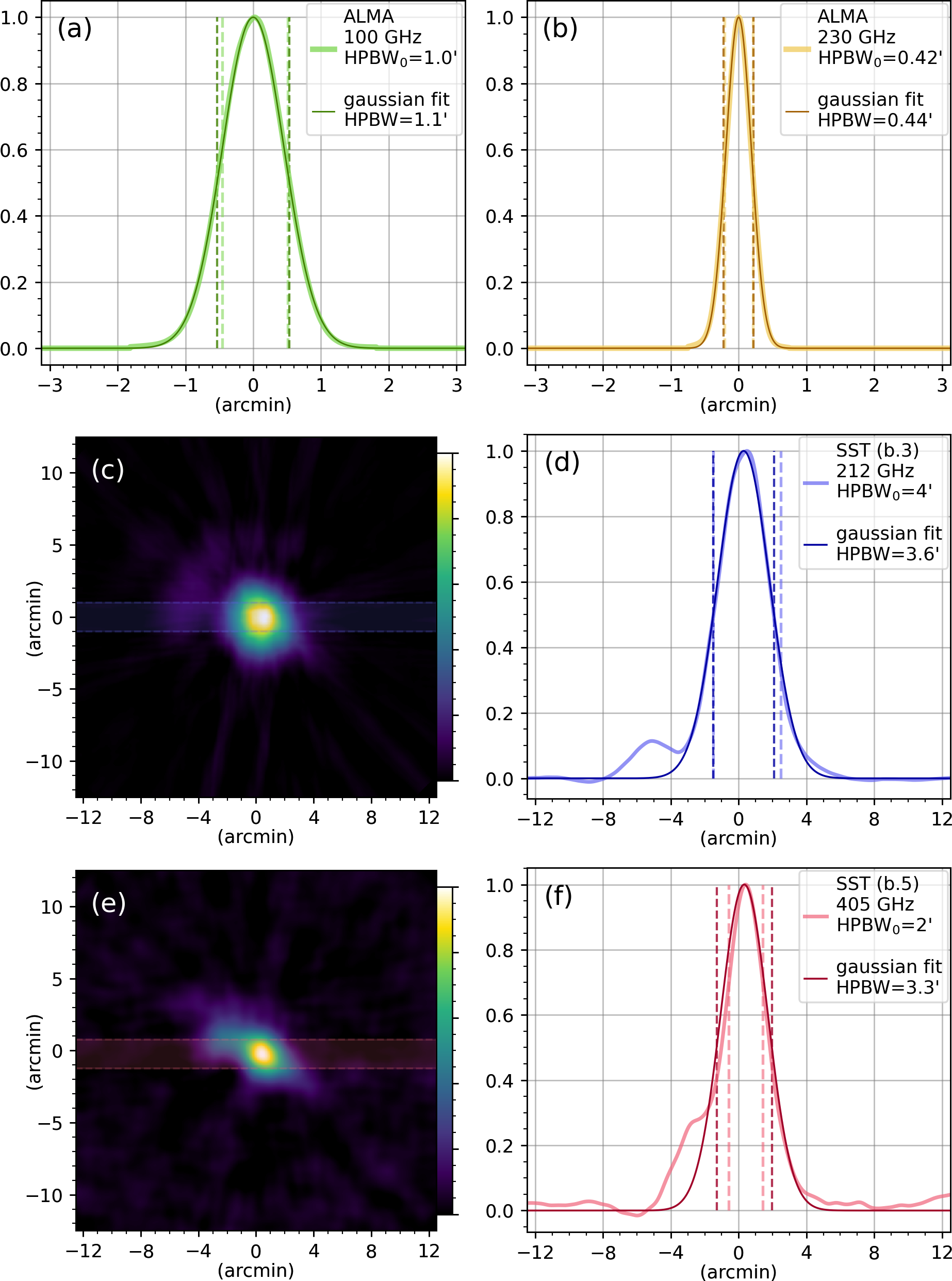}
\caption{Reconstructed 1-D beam profiles from ALMA's single-dish at 100 (a) and 230 GHz (b), and from SST's antenna at 212 (d) and 405 GHz (f). Thicker lighter curves represent reconstructed beam profiles (\BW$_0$ ), and thinner darker curves represent Gaussian fits of the 1-D profiles. Panels c (beam \#3, 212 GHz) and e (beam \#5, 405 GHz) show 2-D SST beam profiles interpolated in 600$\times$600 pixels, \ang{;;2.4}/pixel.}
\label{fig:beams} \end{figure*}

One of the most important features that influence the measurement of the radius and limb brightening is the antenna beam width. ALMA's single-dish nominal half-power beam widths, \hpN, are \ang{;;58} at 100 GHz and \ang{;;25} at 230 GHz. SST's antenna nominal \hpN\ are \ang{;4;} and \ang{;2;} at 212 and 405 GHz, respectively. However, to compare the radius measurement methods and to perform solar scan simulations, the true antenna beam profiles are necessary.

ALMA beams were reconstructed using the derivatives of the solar equator \tb\ profiles extracted from all the solar maps, such as those of Fig.~\ref{fig:maps}. Limb brightening/darkening affects the derivative, however due to the number of maps (the derivatives are superimposed), and to the high spatial resolution of ALMA single-dish, 
\tb\ profiles derivatives are good approximations for the beam. Differences between this approximation and a 2D Gaussian beam model reconstructed with a point source, are negligible in this study. The average of all derivatives is the 1-D reconstructed beam, which are shown in panels \textit{a} and \textit{b} of Fig.~\ref{fig:beams}.

The 2-D SST beam profiles were determined using a solar scans' tomography procedure described in \citet{costa02}, and are shown in panels \textit{c} (212 GHz) and \textit{e} (405 GHz) of Fig.~\ref{fig:beams}. The SST 1-D beam profiles were obtained by averaging a band of \ang{;2;} width centered on the 2-D profile peak, and are shown in panels \textit{d} and \textit{f} of Fig.~\ref{fig:beams} for 212 and 405 GHz, respectively. As the SST beam is asymmetric, 1-D convolutions may differ from 2-D convolutions. To analyze that, we performed 2-D convolutions of SST beams and synthetic \tb\ profiles, with \lb\ of 10\%, 20\%, 30\% and 40\%, and \Rp\ of \ang{;;960} and \ang{;;970}. The radii and \lb\ derived from the 2-D convolutions are very close to the 1-D convolutions (differences less than \ang{;;0.5} in radius and less than 1\% in \lb). Therefore, we use the 1-D beam for the analysis.

These four 1-D beams at 100, 212, 230, and 415 GHz are shown in panels (a), (b), (d), and (f) of Fig.~\ref{fig:beams}, respectively, were then fit by a Gaussian function to estimate the real HPBW. These fits, also shown in each panel of Fig.~\ref{fig:beams}, yield values of \ang{;1.1;}, \ang{;3.6;}, \ang{;0.44;} and \ang{;3.3;} for 100, 212, 230, and 405 GHz, respectively. Here we use the HPBW obtained from the Gaussian fits as the reference HPBW.

\section{Radius Determination} \label{sec:rad_determ}

To determine the sub-THz solar radius, we use two methods widely reported in the literature: the \ip\ and the \hp. The \ip\ defines the radius as the region where the limb profile has an inflection-point; we note that the limb is above the $\tau=1$ level at optical wavelengths \citep{thuillier11}. In the \hp, the radius is taken as the region where the map intensity is half that of the quiet-Sun level. Our goal was to investigate which method is more biased toward beam profiles and the presence of limb brightening.

\begin{figure*}
\includegraphics[width=1.4\columnwidth]{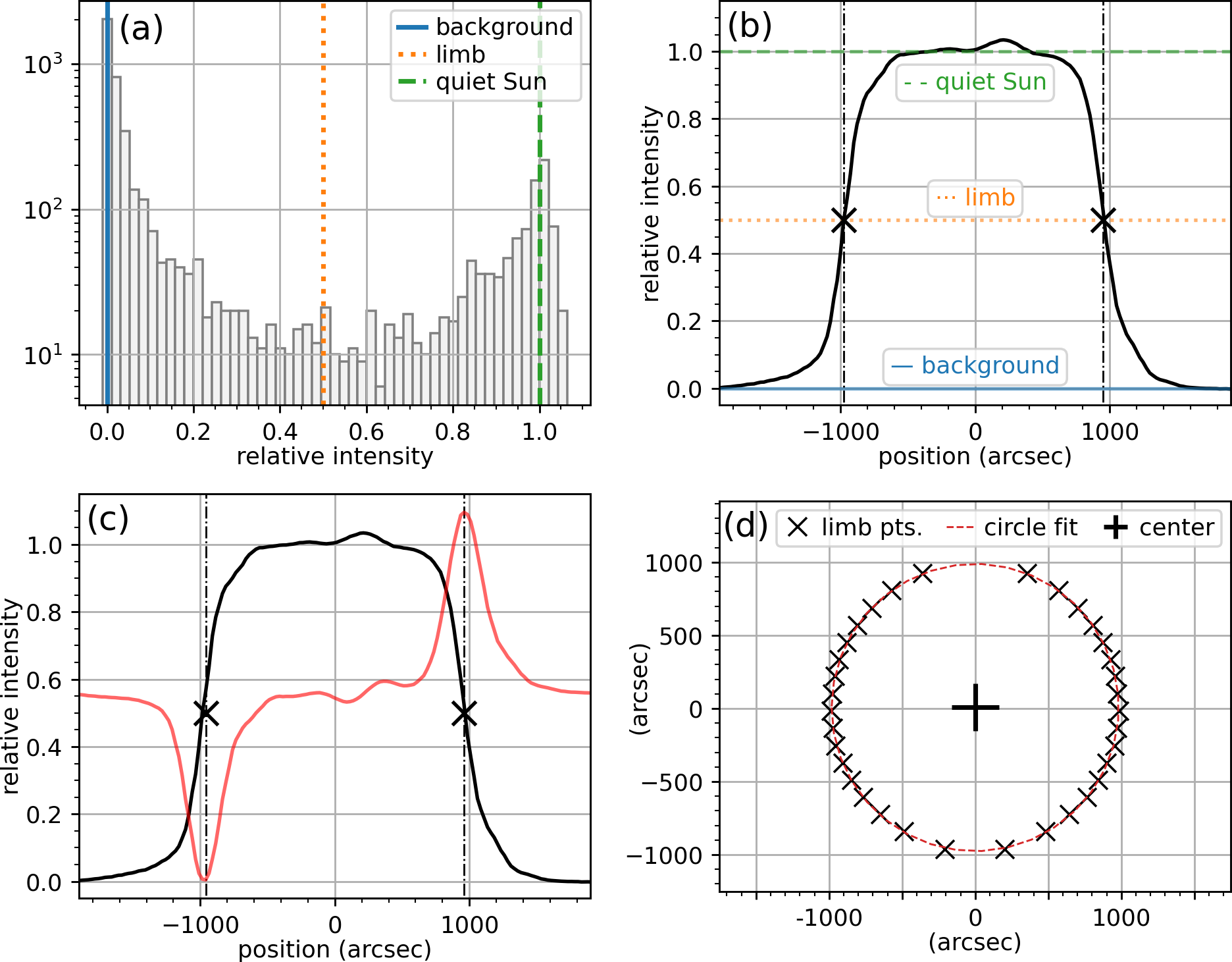}
\caption{Steps of solar radius measurement. Top panels show the \tb\ values distribution of one map (a) and limb points coordinates determination of a single scan by the \hp (b), where background level corresponds to the solid blue line, limb level corresponds to the dotted orange line, and quiet-Sun level corresponds to the dashed green line; limb points coordinates determination by the \ip (c), where minimum and maximum values of the numerical differentiation of a scan (red curve) correspond to the limb coordinates (black crosses); limb coordinates (black crosses) extracted from a solar map used to fit a circle (red dashed line) and determine the solar radius (d). See also Fig. 2 in \citet{menezes21}.}
\label{fig:method} \end{figure*}

All solar maps were corrected accordingly for the eccentricity of the Earth's orbit that makes the apparent solar radii vary over the course of a year. After applying this correction, the first step for both methods is to extract the solar limb coordinates from each map. For the \ip, the limb coordinates (black $\times$ in panels \textit{c} and \textit{d} of Fig.~\ref{fig:method}) are defined as the maximum and minimum points of the numerical derivative (red curve in panel \textit{c}, Fig.~\ref{fig:method}) of each scan across the solar disk.

In the \hp, the solar limb is located at the mean \tb\ value between the sky background level and the quiet-Sun level, \tqs. The background level is defined as the most common \tb\ value, or mode, outside the solar disk (solid blue lines in panels \textit{a} and \textit{b} of Fig.~\ref{fig:method}) and the \tqs value is the \tb\ median value in the center of the solar disk (green dashed lines in panels \textit{a} and \textit{b} of Fig.~\ref{fig:method}) constrained within a region of radius of \ang {;7.5;} (blue circle in Fig.~\ref{fig:maps}b). The interpolated points corresponding to the limb level (orange dotted lines in panels \textit{a} and \textit{b} of Fig.~\ref{fig:method}) are the limb coordinates (black $\times$ in panels \textit{b} and \textit{d} of Fig.~\ref{fig:method}).

These extracted limb points are then fit by a circle (red dashed line in panel \textit{d} of Fig.~\ref{fig:method}) using a least-squares method to determine the center coordinates. The radius, \Rt, of each map is calculated as the average of the center-to-limb distances. 

As described by \cite{menezes21}, some criteria are adopted to avoid extracting limb points associated with active regions and instrumental or random observational errors introduced by high Earth's atmospheric attenuation. For the limb point extraction, only those with a center-to-limb distance between \ang{;;815} ($0.85$ \Rt) and \ang{;;1100} ($1.15$ \Rt) are considered. During the radius calculation, successive circle fits are made. For each fit, the points with center-to-limb distance, $r$, outside the interval $\bar{R}-\ang{;;10}<r<\bar{R}+\ang{;;10}$ are discarded, and then a new circle fit is performed with the remaining points. This process is repeated until no other point is discarded. If there are less than 10 points (out of the usual 34) remaining, the entire map is discarded; otherwise, the radius is calculated. If the standard deviation is below \ang{;;20}, then the calculated radius is stored and the next map is submitted to this process. The same method was applied to the data from both telescopes.



Over 3000 solar maps  observed within the period of 2015-2018 were analyzed and the results of the radius determination for each frequency are listed on Table~\ref{tab:radii} for both the \hp\ and the \ip. Also listed on the last two columns of Table~\ref{tab:radii} are the predicted values of the solar radius, \Rssc, and the respective convolved values, \RsscC, using the SSC atmospheric model \citep{selhor05}, further explained in the next section.

The solar radius results (Table~\ref{tab:radii}) indicated that the \hp\ yield larger radius values when compared to the \ip\ -- the values derived from ALMA maps are about \ang{;;1} larger, while the ones derived from SST maps are about \ang{;;10} larger. Also, the \ip\ radii are closer to the radius measurements reported in the literature \citep{selhor19b, alissan20, menezes21}. 
Moreover, considering the uncertainties, the radii are in agreement with the SSC model radius predictions, \Rssc\ and \RsscC, except \Rssc\ at 100 GHz and \RsscC\ at 212 GHz.

\begin{table}
\centering
\caption{Average radii at sub-THz frequencies determined using the \hp\ and the \ip, and radii derived from SSC model.}
\begin{tabular}{lcccc}
\hline
Frequency & \hs Half-power \hs & \mc{3}{c}{Inflection-point \hs} \\
(GHz) & \Rb\ ($''$) & \Rb\ ($''$) & \Rssc\ ($''$) & \RsscC\ ($''$) \\
\hline
100 (\alma) &   $969\pm2$   & $968\pm3$ &  964.2  &  966.4 \\
212 (\sst)  &   $967\pm3$   & $963\pm3$ &  963.6  &  969.0 \\
230 (\alma) & $964.1\pm1.7$ & $963\pm2$ &  963.5  &  965.0 \\
405 (\sst)  &   $965\pm5$   & $963\pm5$ &  962.9  &  967.4 \\
\hline

\end{tabular} \label{tab:radii} \end{table}

\section{Limb brightening}  \label{sec:LB}

\subsection{Simulated \textit{T\textsubscript{B}} profiles}

\begin{figure*}
\includegraphics[width=1.4\columnwidth]{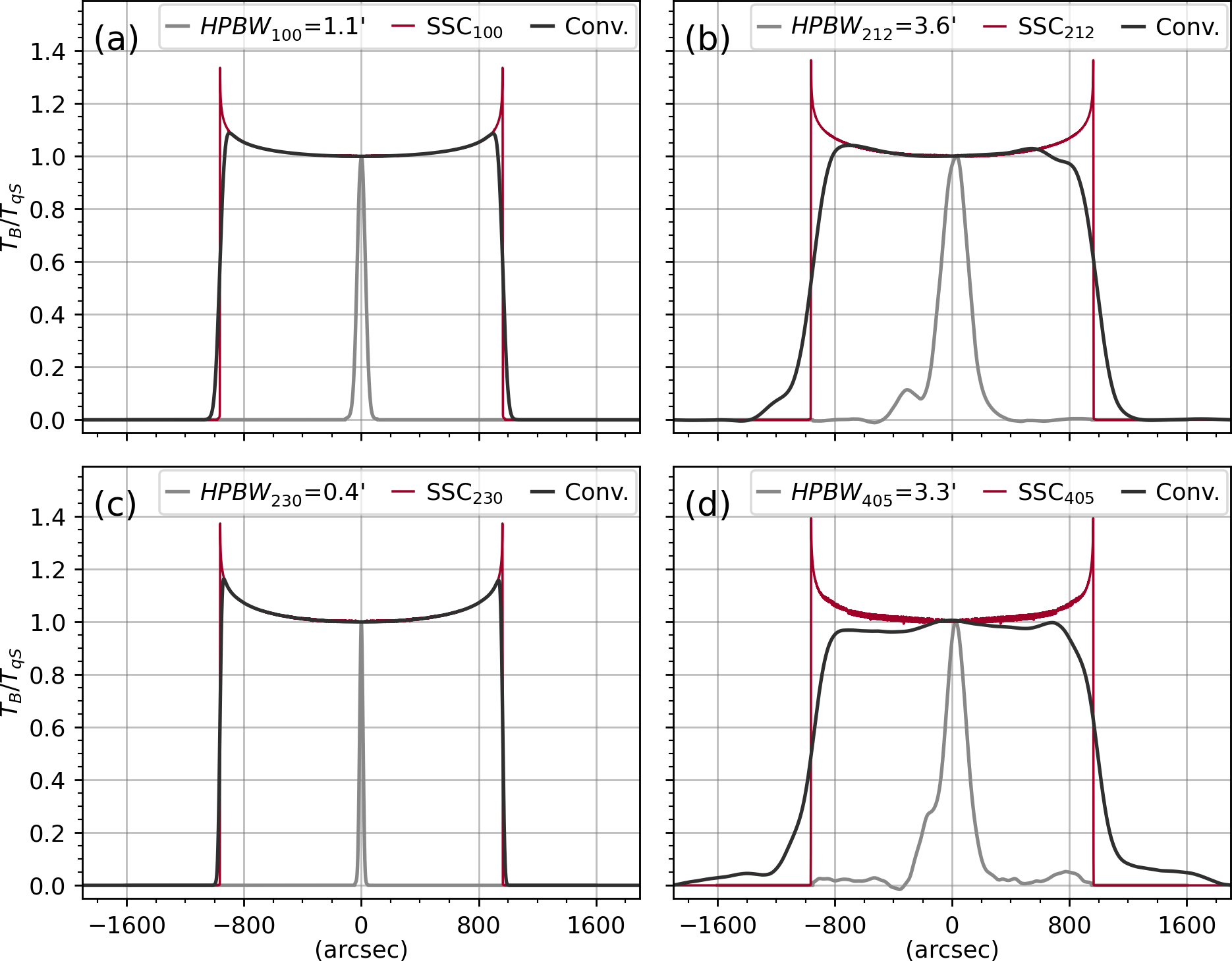}
\caption{Convolved \tb\ profiles (black lines) from the convolution of the telescope beams (gray lines) and the SSC \tb\ profiles (red lines) at 100 (a), 212 (b), 230 (c), and 405 GHz (d).}
\label{fig:conv} \end{figure*}

Atmospheric solar models, such as the SSC model \citep{selhor05} -- a 2-D solar atmospheric model, which takes into account the curvature of the Sun and may include spicules -- predict limb brightening of the solar disk at sub-THz frequencies. High \lb\ of the solar disk  and the antenna beam shape can affect the determination of the radius that could be overestimated, while \lb\ could be underestimated. To investigate this, we simulated single scans over the solar disk, which consist of 1-D convolutions of the beam profiles with the synthetic \tb\ profiles. The synthetic profiles are generated by scaling SSC profiles with limb brightening levels above quiet-Sun level, \lbp, from 0 to 40\% (in 5\% steps), and radius, \Rp, from \ang{;;960} to \ang{;;976} (varying by \ang{;;1}), resulting in a total of 153 \tb\ profiles. Examples of the SSC \tb\ profiles (red curves) convolved with SST and ALMA's beam profiles (gray curves) are shown in Fig.~\ref{fig:conv} as a black curve.

After the convolution, the radius, \Rc, of the convolved \tb\ profile is calculated both by the \hp\ and by the \ip. Then \Rp\ is subtracted from \Rc\ to determine the difference, \dr, the height above the photosphere. By varying \Rp\ for the same \lbp\ value, \dr\ remains with the same value; also, for \lbp = 0\%, \dr\ is negligible. In addition to the radius, the limb brightening level, \lbc, of the convolved profiles are also determined in order to assess the \lb\ decrease after the beam convolution.

Next we estimate a possible range of \lb\ for each average radius, \Rb. First, for each \Rp, we select the four \Rc\ closest to \Rb. From these four values, only those in the range \Rb$-\sigma\leq\Rc\leq$\Rb$+\sigma$ are considered, where $\sigma$ is the standard deviation of \Rb\ for each frequency. Then, we select the \lb\ of the \tb\ profiles that corresponds to these valid \Rc. Moreover, two linear fits are performed: one with the lowest \lb\ of each \Rp, and another onefor the highest \lb\ of each \Rp, resulting in a \lb\ band that ranges from 0\% to 40\%, as a function of \Rp. Finally, we select the lower and the upper value of \lb\ band for the \Rb\ of the solar disk, thus estimating a \lb\ range.

\subsection{Limb brightening estimate}

\begin{figure}
\includegraphics[trim=10pt 10pt 10pt 10pt, clip=true, width=1.0\columnwidth]{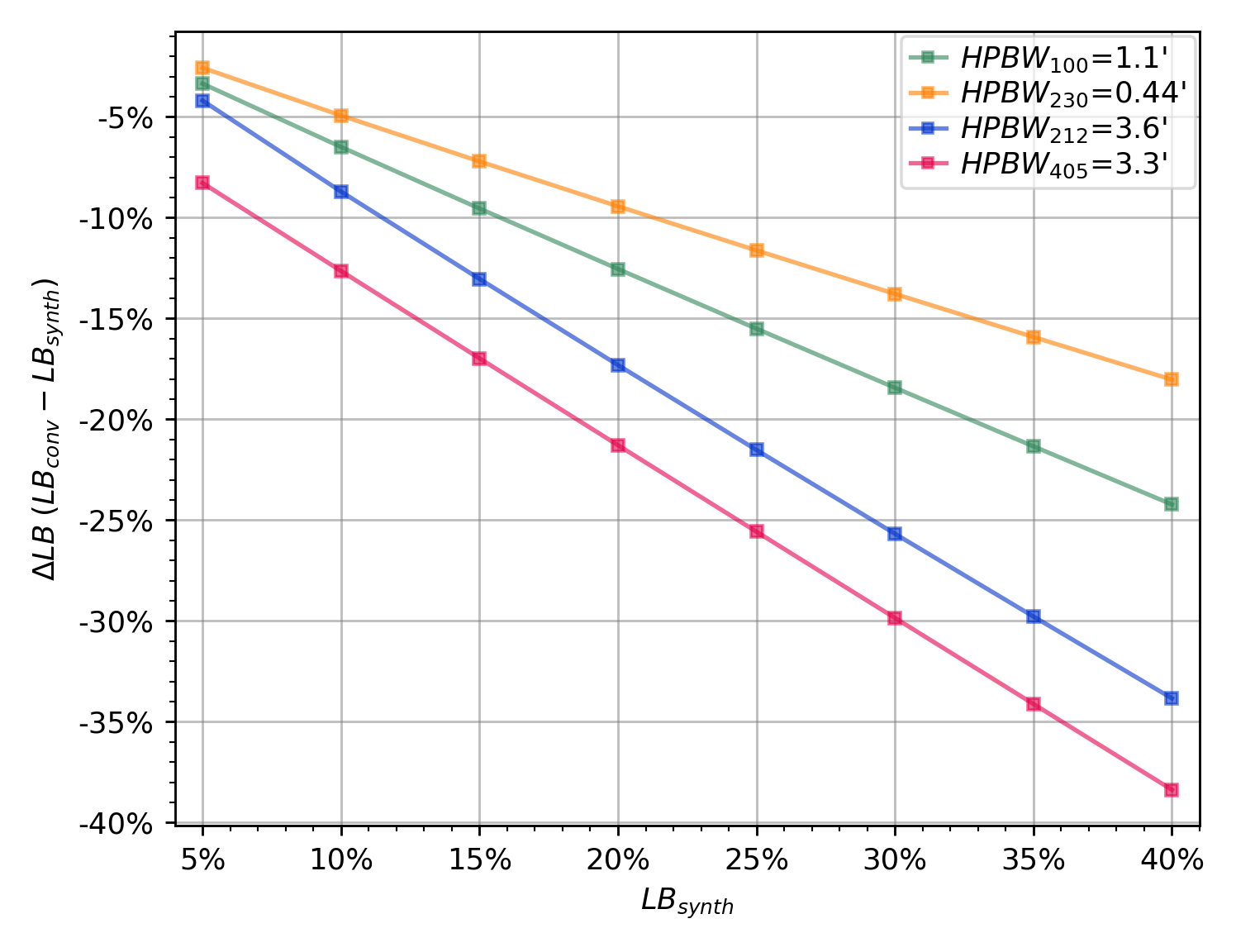}
\caption{Limb brightening decrease, \Dlb\ (\lbp\ -- \lbc), for the respective \lbp, at 100 GHz (green), 230 GHz (yellow), 212 GHz (blue) and 405 GHz (red).}
\label{fig:LBvsLB} \end{figure}

In addition to the radius, \lb\ is also affected by the shape of the antenna beam. In Fig.~\ref{fig:conv} and Table~\ref{tab:LB}, the \lb\ obtained from the SSC \tb\ profiles are much higher than \lbc\ of the respectively convolved profiles. In the case of the \tb\ profiles at 212 and 405 GHz, \lbc\ is very low and the convolved profiles are asymmetric, which are a result of wider beams with high secondary lobes. The decrease in \lb\ can also be seen in Table~\ref{tab:LB} that lists \lbc, and in Fig.~\ref{fig:LBvsLB} that shows the decrease, \Dlb\ (\lbp\ -- \lbc), for the respective \lbp.

\begin{figure*}
\includegraphics[width=1.4\columnwidth]{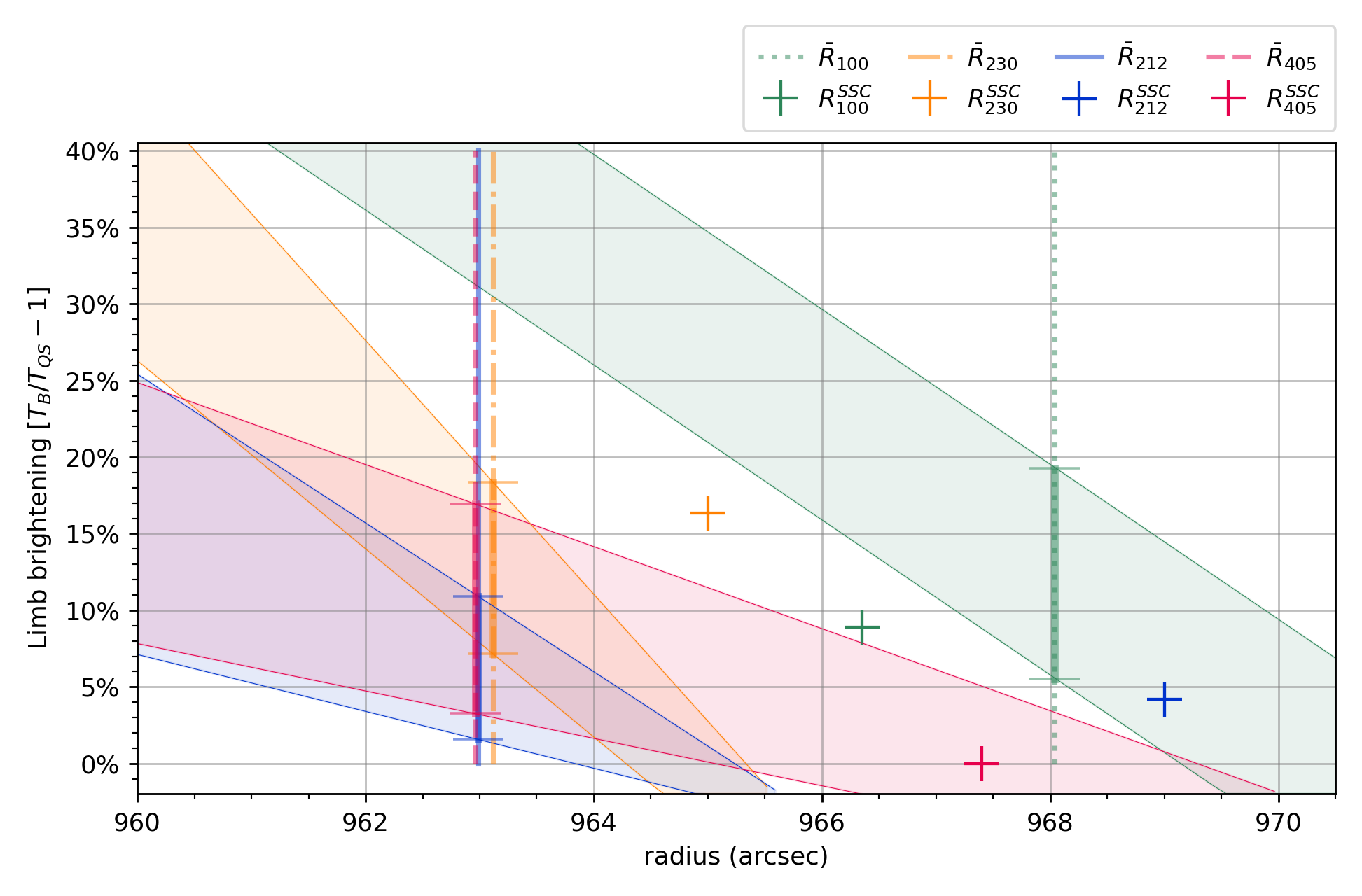}
\caption{Limb brightening, \lb, estimate bands (semitransparent colors) as a function of solar radius derived from the \ip. The crosses are the model predicted radii, \Rssc, and limb brightening levels, \lb, derived from the SSC \tb\ profiles convolved with the beams. The vertical lines are the observed average radii, \Rb, at 100 (dotted green line), 212 (solid blue line), 230 (dashed-dotted yellow line) and 405 GHz (dashed red line). Each frequency is represented by the colors green (100 GHz), blue (212 GHz), yellow (100 GHz) and red (405 GHz).}
\label{fig:LBvsRp} \end{figure*}

We use the \Rc\ calculated by the \ip\ to estimate a band of possible \lb\ for each frequency, by performing several solar scan simulations. The bands of \lb\ as a function of the solar radius, \Rp, are shown in Fig.~\ref{fig:LBvsRp}. For example, the \lb\ at 230 GHz has an upper limit that ranges from 40\% to 0\% for \Rp\ in the interval from $\sim$\ang{;;960.5} to $\sim$\ang{;;965.5}; whereas the lower limit ranges from 25\% to 0\% for a \Rp\ range within \ang{;;960} to $\sim$\ang{;;964.2}. Therefore, based on the lower and upper limits and on \Rb\ (average radii), \lb\ estimate at 230 GHz ranges from 6.4\% to 17.6\%.

\begin{table}
\centering
\caption{Limb brightening levels derived from SSC model, and limb brightening estimates based on the average radii.}
\begin{tabular}{llclcc}
\hline
Frequency & \hs & \lb\ for \Rb & \hs & \mc{2}{c}{SSC model} \\
(GHz) & & (\%) & & \lb\ (\%) & \lbc(\%)\\
\hline
100 (\alma) &  & 5.0 - 18.7 &  & 33.6 &  8.9 \\
212 (\sst)  &  & 1.6 - 12.0 &  & 36.5 &  4.2 \\
230 (\alma) &  & 6.4 - 17.6 &  & 37.4 & 16.3 \\
405 (\sst)  &  & 3.3 - 17.0 &  & 39.3 &  3.2 \\
\hline
\end{tabular} \label{tab:LB} \end{table}

In Table~\ref{tab:LB}, \lb\ estimate ranges based on \Rb\ (average radii) are listed. The \lb\ ranges at 100 and 230 GHz agree with the values reported by \citet{selhor19}. At 212 and 405 GHz, there are no measured values, because the SST beams are too wide and, hence, no \lb\ are seen in the solar maps. However, considering the proximity in frequency to 230 GHz, the \lb\ range at 212 GHz contains the \lb\ at 230 GHz reported by \citet{selhor19}.

The \lb\ ranges estimate do not agree with the \lb\ derived from SSC's \tb\ profiles (second last column of Table~\ref{tab:LB}), which predicts a much higher \lb. Nevertheless, \lbc\ (last column of Table~\ref{tab:LB}) derived from the convolved SSC model \tb\ profiles are within the estimate ranges, except at 405 GHz, that shows a lower \lbc\, albeit very close to the lower limit. As stated in Section~\ref{sec:beam}, SST beams asymmetyr and large FWHD  could explain such low \lb. Also, SSC profiles \lbc\ at 100 and 230 GHz are respectively lower, however close to the 10.5\% and 17.8\% levels reported by \citet{selhor19}.

\subsection{Limb brightening \textit{versus} radius determination method}

\begin{figure}
\includegraphics[trim=10pt 10pt 10pt 10pt, clip=true, width=1.0\columnwidth]{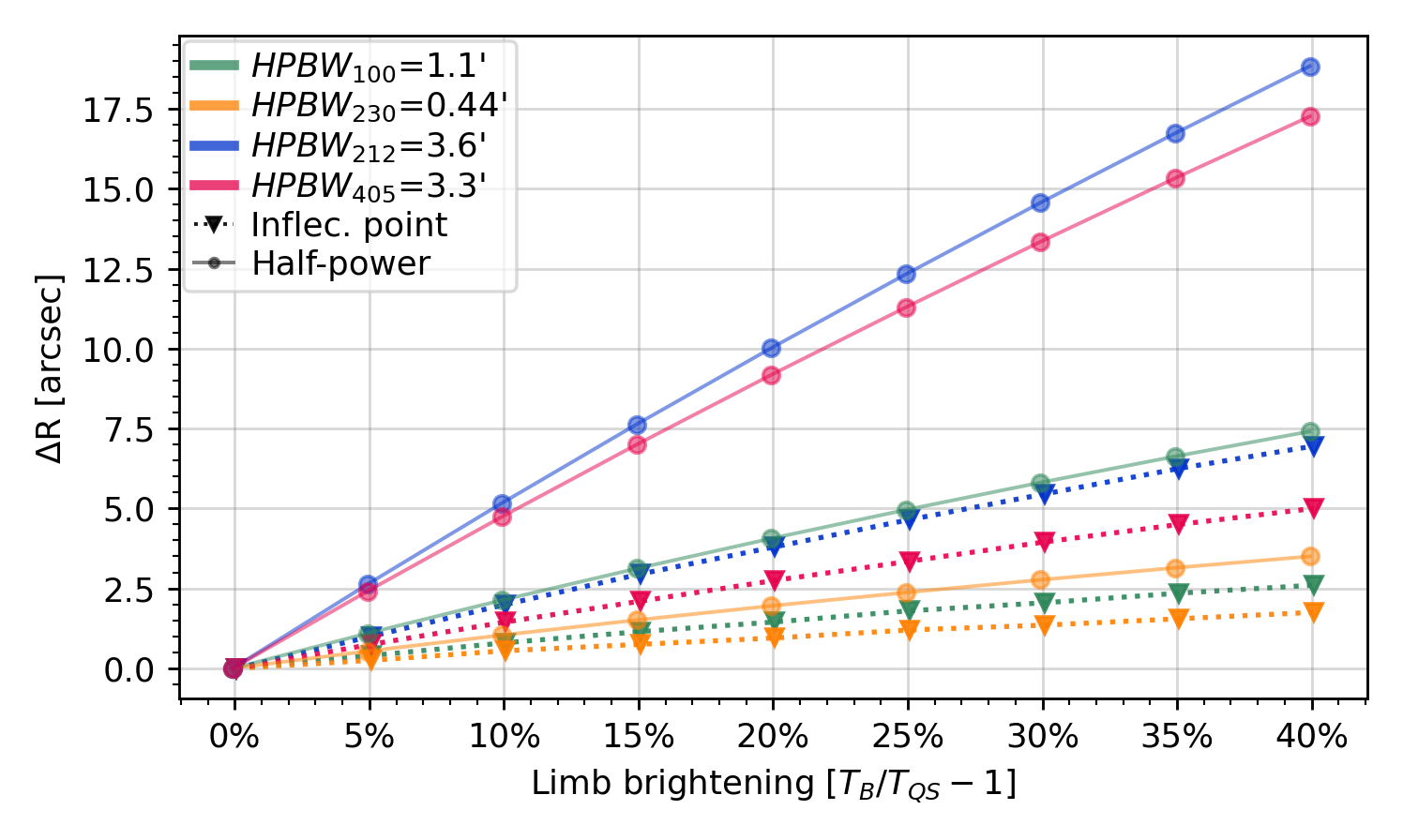}
\includegraphics[trim=0pt 10pt 10pt 10pt, clip=true, width=1.0\columnwidth]{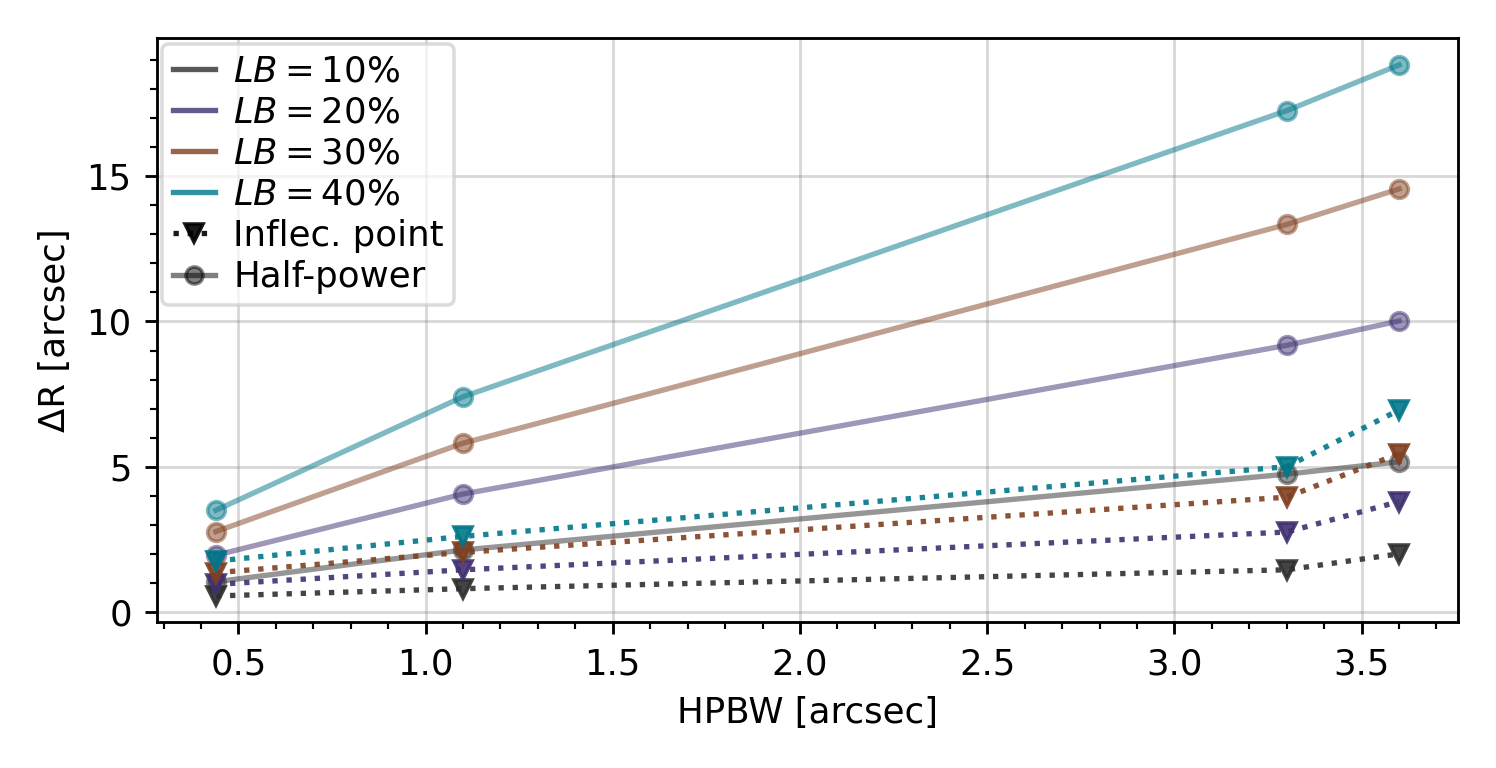}
\caption{Radius increases, \dr, calculated as the subtraction, $\Rc-\Rp$, of the synthetic profile radius from the convolved profiles. 
{Top panel}: \dr\ for each beam profile is represented by green (100 GHz), yellow (230 GHz), blue (212 GHz) and red (405 GHz) lines. {Bottom panel}: \dr\ for each limb brightening level, \lb, is represented by black (\lb~=~10\%), purple (20\%), brown (30\%) and turquoise (40\%) lines. In both panels, the solid color lines with circles represent \dr\ derived from the \hp, whereas the dotted color lines with inverted triangles represent \dr\ derived from the \ip.}
\label{fig:DRvsLB} \end{figure}

The discrepancies between both methods are more evident in Table~\ref{tab:conv} and in Fig.~\ref{fig:DRvsLB}. Values of \Rc\ and \lbc\ of the convolved profiles are listed in Table~\ref{tab:conv}. Fig.~\ref{fig:DRvsLB} shows the radius increases, \dr, \textit{i. e.} the difference between the radii of the \tb\ profiles, \Rp, and the convolved profiles, \Rc. In the top panel, \dr\ is shown as a function of \lb, and in the bottom panel, as a function of \BW. This confirms that the measured radii are affected by the shape of the beam and \tb\ profiles.

\begin{table*}
\centering
\caption{Values of \lbc\ and \Rc\ derived from 1-D convolutions of beam profiles and synthetic \tb\ profiles.}
\begin{tabular}{lcccc}
\hline
Beam          & \lbp\   & \lbc    & Half-power  & Inflection-point \\
              & (\tbqs) & (\tbqs) & \Rc\ ($''$) & \Rc\ ($''$) \\
\hline
100 GHz             &  5\%  &  1.6\%  &  965.4  &  964.7 \\
\BW=\ang{;1.1;}     & 10\%  &  3.5\%  &  966.4  &  965.0 \\
\Rp=\ang{;;964.2}   & 15\%  &  5.5\%  &  967.4  &  965.4 \\
                    & 20\%  &  7.5\%  &  968.3  &  965.7 \\
                    & 25\%  &  9.5\%  &  969.2  &  966.0 \\
                    & 30\%  & 11.6\%  &  970.1  &  966.3 \\
                    & 35\%  & 13.7\%  &  970.9  &  966.6 \\
                    & 40\%  & 15.8\%  &  971.7  &  966.8 \\
\hline
212 GHz             &  5\%  &  0.8\%  &  966.3  &  964.7 \\
\BW=\ang{;3.6;}     & 10\%  &  1.3\%  &  968.8  &  965.7 \\
\Rp=\ang{;;963.6}   & 15\%  &  2.0\%  &  971.3  &  966.6 \\
                    & 20\%  &  2.7\%  &  973.7  &  967.5 \\
                    & 25\%  &  3.5\%  &  976.0  &  968.3 \\
                    & 30\%  &  4.3\%  &  978.2  &  969.1 \\
                    & 35\%  &  5.2\%  &  980.4  &  969.9 \\
                    & 40\%  &  6.2\%  &  982.5  &  970.6 \\
\hline
230 GHz             &  5\%  &  2.4\%  &  964.1  &  963.8 \\
\BW=\ang{;0.44;}    & 10\%  &  5.1\%  &  964.6  &  964.1 \\
\Rp=\ang{;;963.5}   & 15\%  &  7.8\%  &  965.1  &  964.3 \\
                    & 20\%  & 10.6\%  &  965.5  &  964.5 \\
                    & 25\%  & 13.4\%  &  966.0  &  964.8 \\
                    & 30\%  & 16.2\%  &  966.4  &  964.9 \\
                    & 35\%  & 19.1\%  &  966.7  &  965.1 \\
                    & 40\%  & 22.0\%  &  967.1  &  965.3 \\
\hline
405 GHz             &  5\%  & -3.3\%  &  965.5  &  963.7 \\
\BW=\ang{;3.3;}     & 10\%  & -2.6\%  &  967.8  &  964.4 \\
\Rp=\ang{;;962.9}   & 15\%  & -2.0\%  &  970.1  &  965.0 \\
                    & 20\%  & -1.3\%  &  972.2  &  965.7 \\
                    & 25\%  & -0.6\%  &  974.4  &  966.3 \\
                    & 30\%  &  0.1\%  &  976.4  &  966.9 \\
                    & 35\%  &  0.9\%  &  978.4  &  967.5 \\
                    & 40\%  &  1.6\%  &  980.3  &  968.0 \\
\hline
\end{tabular} \label{tab:conv} \end{table*}

In both methods, 
\dr\ increases with both \lb\ and \BW. However, \dr\ is larger when the \hp\ is used, being less pronounced for narrow \BW\ and more pronounced for wider \BW. This means that the \ip\ is less susceptible to the shape of the beam and \tb\ profiles. Therefore, the \ip\ carries less bias to the calculation of the solar radius and should thus be preferred for radius determination.

\section{Conclusions} \label{sec:conclusion}

In this work we determine the average radius of the full solar disk at 100, 212, 230, and 405 GHz, using data from SST and ALMA's single-dish observations. We compare the \hp\ and the \ip\ to find which one is more suitable for this kind of measurement. By performing solar scans simulations (1-D convolutions), we identify how the \lb\ in \tb\ profiles and the \BW\ of the antennas inflict larger bias in the measurements. Moreover, we use the simulation output to estimate ranges of probable \lb\ for solar \tb\ profiles observed at sub-THz frequencies.

The solar scan simulation estimates of \dr\ indicated that the combination of solar limb brightening in association with the radio-telescope beam width and shape can affect the radius determination depending on the method used. We found that the \ip\ is less prone to the irregularities of the telescope beam and to the variations of the \lb\ profile of the Sun. In conclusion, the \ip\ results in a lower bias in solar radius determination and, therefore, is more suitable for such task.

The higher precision solar radii calculation allowed us to investigate the presence of limb brightening and estimate \lb\ of the Sun at sub-THz frequencies. This is done by simulating \tb\ profiles with varying \lb\ and convolving them with the measured beam profiles. The limb brightening determination from solar scan simulations yields ranges of \lb\ that agree with previous measurements \citep[100 and 230 GHz,][]{selhor19} and is able to estimate \lb\ from low spatial resolution solar maps such as those from SST. However, the values are not in agreement with the \tb\ profiles generated by the SSC model.

As stated by \citet{selhor19}, some observations reported brightening values much smaller than the expected values predicted by the models, with discrepancies being particularly large at millimeter wavelengths. Those discrepancies could be due to the presence of chromospheric features such as spicules located close to the limb or, in the case of SST maps, probably due to the large beam width and deformations. In fact, the model represents average approximations, which do not always reproduce the observations.

Measurements of the Sun at radio frequencies with high precision can be difficult, since it depends directly on high spatial resolution solar maps. In spite of that, we were able to determine the solar radius and estimate the limb brightening at sub-THz frequencies. Our results are crucial to test and improve solar atmospheric models, and better understand the solar atmosphere, since they probe directly different low lying atmospheric layers. More studies of this kind at other radio frequencies are needed to achieve such objectives.

\section*{Acknowledgements} \label{sec:acknow}

This article has been accepted for publication in Monthly Notices of the Royal Astronomical Society Published by Oxford University Press on behalf of the Royal Astronomical Society.

We thank the referee, Dr. C. Alissandrakis, for careful consideration of the manuscript and comments that helped its improvement. We acknowledge the financial support for operation of the Solar Submillimeter Telescope (SST) from S{\~a}o Paulo Research Foundation (FAPESP) Proc. \#2013/24155-3 and AFOSR Grant \#FA9550-16-1-0072. F. M. thanks MackPesquisa and CAPES for the scholarship. A. V. acknowledges partial financial support from the FAPESP, grant number \#2013/10559-5.
C.L.S. acknowledges financial support from the FAPESP, grant number \#2019/03301-8. C.G.G.C. is grateful to CNPq (grant 307722/2019-8) and CAPESPRINT (1035531P) for providing support to this research.

This work is based on data acquired at {Complejo Astronómico El Leoncito}, operated under agreement between the {Consejo Nacional de Investigaciones Cient\'ificas y T\'ecnicas de la Rep\'ublica Argentina} and the {National Universities} of {La Plata}, {C\'ordoba} and {San Juan}.

This work makes use of the following ALMA data: ADS/JAO.ALMA \#2011.0.00020.SV, \#2016.1.00050.S, \#2016.1.00070.S, \#2016.1.00156.S, \#2016.1.00166.S, \#2016.1.00182.S, \#2016.1.00201.S, \#2016.1.00202.S, \#2016.1.00423.S, \#2016.1.00572.S, \#2016.1.00788.S, \#2016.1.01129.S, \#2016.1.01408.S, \#2016.1.01532.S, \#2017.1.00009.S, \#2017.1.00870.S, \#2017.1.01138.S. ALMA is a partnership of ESO (representing its member states), NSF (USA) and NINS (Japan), together with NRC (Canada), MOST and ASIAA (Taiwan), and KASI (Republic of Korea), in cooperation with the Republic of Chile. The Joint ALMA Observatory is operated by ESO, AUI/NRAO and NAOJ.

\section*{Data Availability}

The data from the Solar Submillimeter-wave Telescope (SST) underlying this article were provided by the Center for Radio Astronomy and Astrophysics Mackenzie (CRAAM) by permission. Data will be shared upon request to the corresponding author with permission of CRAAM.

The data from the Atacama Large Millimeter/submillimeter Array (ALMA) underlying this article are available in ALMA Science Verification Data page at \texttt{https://almascience.eso.org/alma-data/science-verification} (ADS/JAO.ALMA \#2011.0.00020.SV), and in ALMA Science Archive page at \texttt{http://almascience.eso.org/asax/} (the rest of ALMA data).


\bibliographystyle{mnras}
\bibliography{Main.bib} 


\bsp	
\label{lastpage}
\end{document}